# Biological Effects of Stellar Collapse Neutrinos

J. I. Collar

Department of Physics and Astronomy
University of South Carolina
Columbia, South Carolina 29208, USA

electronic address: ji.collar@sc.edu

astro-ph/9505028   1 Dec 1995

## Abstract

Massive stars in their final stages of collapse radiate most of their binding energy in the form of MeV neutrinos. The recoil atoms that they produce in elastic scattering off nuclei in organic tissue create radiation damage which is highly effective in the production of irreparable DNA harm, leading to cellular mutation, neoplasia and oncogenesis. Using a conventional model of the galaxy and of the collapse mechanism, the periodicity of nearby stellar collapses and the radiation dose are calculated. The possible contribution of this process to the paleontological record of mass extinctions is examined.

Several theories intend to explain the record of biological extinctions that sprinkle the history of the Earth. They fall into two groups: terrestrial phenomena (e.g., tectonically induced climatic changes and massive volcanic eruptions) and extraterrestrial processes such as variations in the solar constant or large meteorite impacts. Detailed accounts of these approaches can be found in Refs. [1-3].

A proposal that ionizing radiation from nearby supernova explosions may have had catastrophic effects on the Earth's ozone layer [4] has been revisited recently [5]. While this may have caused one of the "Big Five" major extinctions, the estimated nearby supernovae rate is insufficient for these to be responsible for the bulk of extinctions in the last 600 Myr. The apparent periodicity of extinction episodes is intensity-dependent; the disappearance of ~ 65% (~ 30%) of living species occurs every ~ 100 Myr (~ 30 Myr) [1-3,6].

It is a common and erroneous assumption to identify stellar collapse with supernova production. As discussed by Bahcall [7], the collapse rate exceeds by a yet unknown but possibly large factor the supernova rate in our galaxy, which is calculated to be between 0.1 and 0.01 $yr^{-1}$. Stellar collapses may lack the optical emission of supernovae, being much more inconspicuous while producing an intense neutrino radiation. Collapse models generally have difficulty in making stars explode, but predict neutrino fluxes that carry away most of the binding energy released. Using the known distribution of stars in the Galaxy and their rate of evolution, Bahcall has estimated the period of stellar collapses in our galaxy to be $\tau_{total}$=11.1 yr [7]. Could these "silent" stellar collapses affect life on Earth, even when the main form of emitted radiation is weakly interacting? The expected extreme Relative Biological Effectiveness (RBE, a commonly used factor that indicates the specific biological effectiveness of any radiation) of neutrino-induced recoils in generating cell transformation and radiocarcinogenesis may indeed guarantee this.

Essential for the determination of the radiation dose imparted by this process is the neutrino fluence, i.e., the number of neutrinos per unit area received at a distance r during the collapse pulse (< 10s), which is $F(r) \cong 6 \cdot 10^{11} (r/10 \text{ kpc})^{-2}$ $\nu/cm^2$ [7]. The energy spectrum of collapse neutrinos is given by a Boltzmann distribution [7] $dF/dE_\nu = \left( F(r) E_\nu^2 / 2T_\nu^3 \right) \exp(-E_\nu/T_\nu)$, where $T_\nu$= 5 MeV for $\nu_e, \bar{\nu}_e$ and 10 MeV for $\nu_{\mu,\tau}, \bar{\nu}_{\mu,\tau}$, extending with appreciable probability up to several tens of MeV. Knowledge of the spatial distribution of stars as a function of their distance r from the Sun is needed to estimate how often one of these "neutrino bombs" goes off nearby the Earth. Here the model of Bahcall and Soneira [7,8] is adopted, where for r < 3 kpc, the fraction of stars in the galactic disk within r is given by $f(r) = 0.0028 \, (r/1 \, \text{kpc})^2$. The expected period of collapses within r is then

$$\tau = \frac{\tau_{total}}{f(r)} = \frac{3.96 \cdot 10^3}{(r/1 \, \text{pc})^2} \, \text{Myr} \quad (r < 3 \text{kpc}), \qquad (1)$$

occuring at a mean distance <r> = (2/3) r. To assume that f(r) has been constant over the last 600 Myr is only a default approximation, since the trajectory of the Sun around the galaxy can only be extrapolated back or forth in time with any degree of precision over a time scale ~ 100 Myr [9].

The most probable interaction process for MeV neutrinos with matter is elastic scattering off nuclei via the neutral current. This is due to an enhancement by a coherence factor roughly proportional to the square of the number of neutrons in the target atom, N, giving a flavor-blind total cross section $\sigma \cong N^2 \, (E_\nu / 10 \, \text{MeV})^2 \, 4 \cdot 10^{-43} \, \text{cm}^2$ and a differential cross section $d\sigma/d\chi \propto (1+\chi)$, where $\chi$ is the cosine of the neutrino scattering angle [10]. The recoil energy of the target nucleus of mass M is given by $T = E_\nu^2 (1-\chi)/(2E_\nu + M)$. The differential rate of recoils induced by a collapse in 1 kg of target matter (containing K nuclei) is obtained from $dR/dT = K \int_{E_{min}}^{E_{max}} (dF/dE_\nu)(d\sigma/dT) \, dE_\nu$. The integral includes all neutrino flavors, $E_{min}$ is the minimal neutrino energy that can contribute to recoils in the interval [T,T+dT] and $E_{max}$ is a suitable upper cut-off to $dF/dE_\nu$. Fig. 1 displays dR / dT for some common constituents of organic tissue. Table 1 shows, for these same elements, the total recoil rate and other relevant quantities (ranges and stopping powers are extracted from the computer code TRIM [12]). Using $C_4H_{40}O_{17}N_1$ as representative of the common elemental proportions in animal tissue, most of the recoils correspond to oxygen (85.5%), followed by carbon (10.8%). With this chemical composition and the $r^{-2}$ scaling law for the fluence, one obtains the mean number of recoils induced in 1 kg of tissue by a collapse within a distance r of the Earth

$$\langle R \rangle \cong \frac{7.71 \cdot 10^5}{(r/1pc)^2} \text{ recoils/kg},  \qquad (2)$$

where the mean distance <r> has been used as the collapse location. Eqs. (1) and (2) determine how often a certain dose of radiation is imparted by the neutrinos; e.g., every ~ 100 Myr a collapse within ~6.3 pc of the Earth will produce ~ $1.9 \cdot 10^4$ recoils / kg in all living tissue. The objective is now to assess the relevance of such an event.

In the last few years, the field of radiation protection has undergone important revisions. The study of how different patterns of spatial and temporal distributions of energy at the subcellular level affect the biological response of cells to very low doses of ionizing radiation has become a very active area of research. This is due to the increasing awareness that high linear-energy-transfer (LET) radiations (alpha particles, fission neutrons, ions) may be responsible for unique biological effects [13-15]. Plainly stated, the (unrestricted) LET is the amount of energy dissipated by a radiation per unit path length.

The nucleus of a cell, specifically the genetic material, is far more radiosensitive than the cytoplasm. Chromosomal change can result from breaks and other types of damage to the DNA chain, leading to mutagenesis. Of special importance are those radiation insults that create irreparable genetic damage without inactivating the cell, which then can become the "founder" cell of an aggregation of mutant or cancerous cells. Not all types of radiation are equally effective in producing important damage at the chromosomal level. Low-LET radiations such as gamma and x-rays disperse their energy over longer distances than their high-LET counterparts, which have characteristic densely-ionizing tracks. There is now a wide consensus that the critical property of radiations at low doses is determined by this spatial pattern of energy deposition over dimensions similar to those of DNA structures (few nm) [14]. For instance, the primary molecular damage of importance in mammalian cells is produced by a localized cluster of atomic interactions overlapping the DNA and producing a multiply damaged chromosomal site [15]. Track structure analysis of high-LET radiations shows energy concentrations at subcellular levels that cannot be at all reproduced by low-LET radiation. Energy depositions of $\gtrsim$ 800 eV within 5-10 nm are unique to alpha particles and heavier ions, totally unattainable for other types of radiation and provoking unrepairable genetic damage [15]. Their frequency of occurrence is not more than 0-4 / cell / Gy (1 Gy, "gray" = unit absorbed dose = $6.24 \cdot 10^{12}$ MeV / kg). A critical question is then whether this type of damage unique to high-LET particles can lead to unique genetic consequences, specifically mutagenic changes and certain types of radiogenic cancer.

The greater effectiveness of the high-LET tracks more than compensates for their smaller number per unit dose. This is illustrated by the microdosimetric concept of Specific Energy, $z = \varepsilon / m$, where $\varepsilon$ is the energy imparted by radiation to matter of mass m [13]. It is a stochastic quantity and depends on the number of energy deposition events in the mass of interest, and therefore on the ionization density of the tracks. Considering a 7 µm diameter sphere as representative of an eukaryotic cell nucleus, single $^{60}$Co events have a mean z of 1.25 mGy, while for fission neutrons z ~ 200 mGy [13]. For an 80 keV oxygen ion, representative of stellar collapse damage (see table), z = 71 mGy. If the region of interest is a DNA segment (2 nm diameter x 2 nm length cylinder), $z = 5.5 \cdot 10^5$ Gy, $22.1 \cdot 10^5$ Gy and $56.0 \cdot 10^5$ Gy, respectively. The utility of z is illustrated as follows: for a neutron exposure (dose) of D= 3 mGy, only D / z = (3 mGy / 200 mGy) = 1.5 % of cell nuclei in the irradiated tissue will receive any damage, while each of them gets the same amount as cells exposed to 200 mGy in a dose-response study [13]. In our case, the exposure can be approximated (using the prevalent Oxygen recoils and Eq. (2)) as $D \cong 0.08$ MeV / recoil $\cdot$ $7.71 \cdot 10^5 (r / 1 \text{ pc})^{-2}$ recoils / kg = $9.88 \cdot 10^{-9} (r / 1 \text{ pc})^{-2}$ Gy, and the fraction of cell nuclei affected is then D / z = D / 71 mGy. Assuming a typical eukaryotic cell diameter of 20 - 40 µm, the number of cells damaged in their nuclei per stellar collapse is:

$$C = \frac{D}{z} \cdot \left( \sim 7 \cdot 10^{10} \frac{\text{cells}}{\text{kg}} \right) = \frac{9.8 \cdot 10^3}{(r / 1 \text{ pc})^2} \frac{\text{cell nuclei}}{\text{kg}}. \quad (3)$$

Given a distribution of spheres of average radius ρ scattered in a medium with volume filling factor φ, the mean number of spheres encountered by a particle is (3/4) φ/ρ per unit length, independent of trajectory. Taking $\sim 3 \cdot 10^7$ nucleosomes per nucleus (adequate for most large animals) and approximating their cylindrical volume to that of a ρ= 4.5 nm sphere, φ~ 0.065. Considering the range of an 80 keV oxygen atom in water, 0.33 µm, this translates into ~ 3.6 nucleosomes damaged per cell nucleus hit, i.e., a

~97% certainty that the cell nuclei affected in Eq. (3) carry important damage at the nucleosome or chromosomal level.

    The recent availability of heavy-ion sources in the field of radiation biology has made it possible to test their possible unique biological effects [16]. Yang and collaborators [17] have irradiated C3H10T1/2 cells employing BEVALAC-ions from carbon to uranium and ranging from 300 to 1000 MeV / u, spanning the LET range 10-1000 keV / μm. A damage irreparability threshold for mammal cells at > 100 keV / μm was established. They found a maximal neoplastic cell transformation rate (malignant cell transformation, the first step in tumor formation) in surviving cells at about 100-200 keV / μm. The LET for the stellar collapse-representative oxygen recoils is precisely 160 - 270 keV / μm for T < 100 keV. More recently, Kadhim et al [18] irradiated murine bone marrow cells with very low doses of 3.3 MeV $^{238}$Pu alpha particles (LET=121 keV / μm) and x-rays (the doses were consistent with single or no tracks crossing the individual cell nuclei). Cells surviving alpha particle irradiation showed a striking transmission of non-clonal radiation-induced aberrations to their clonal descendants, absent in those treated with x-rays. This chromosomal instability in the progeny was still present after 10-13 cell divisions and appeared with a high frequency (40-60%) in cell colonies descendant of those cells surviving the irradiation, a 10% of the total. This suggests a ~ 5% effectiveness in the creation of initiating lesions able to produce the onset of leukemias [18]. A similar effectively-infinite RBE has been suggested for the formation of sister chromatid exchanges in human lymphocytes [19]. These results have been independently confirmed using human dermal fibroblasts and ions ranging from 0.386 - 13.6 MeV / μm, which produced chromosome imbalances and rearrangements like those observed in human solid tumors [20].

    Assuming for the sake of argument that the above ~ 5% effectiveness in producing a possibly malignant cell colony also applies to stellar collapse recoils, Eq. (3) gives the discomforting yield of $490 \cdot (r / 1 \text{ pc})^{-2}$ malignant cell colonies / kg of tissue. This is ~ 4 (12) malignant foci / kg tissue occurring every ~ 30 (100) Myr, an insult that would be severe enough to kill a vast percentage of large animals with a frequency comparable to that of most major extinctions. While smaller living forms may be spared by not enough of their members being directly killed for a non-viable minimal population to be reached, the disappearance of the larger-mass species might influence the ecosystem in such a way that they are also gravely affected.

Several observations are in order. First, the stopping power $S_t$ of an ion displays a broad maximum (at ~ 4 MeV for O in water) and it is therefore possible to have identical values of LET for two very different energies. The nature of the damage induced may however be quite dissimilar. For instance, the high-energy LET-counterpart of an 80 keV oxygen ion in water is at ~ 200 MeV. This is in the energy range generally explored in accelerator experiments, where energy losses are almost entirely due to ionization. The contribution to $S_t$ by direct atomic collisions at lower energies is considerable (see table), and these collisions are of special interest because the energy imparted to an atom is likely to cause disruption of the DNA molecule with a much higher effectiveness than an equivalent delta-electron produced through ionization. This may make neutrino-induced recoils even more effective than other forms of high-LET radiation in producing the dense damage responsible for the special biological effects detected. Fig. 2 displays the results of a computer simulation, where the frequency of the number of atomic collisions with energy above the mean binding energy of the H-O link in water has been calculated (i.e., recoils able to rupture the molecule). This is computed over the volume of a nucleosome and for several types of high-LET radiation. TRIM has been used in obtaining the recoils / ångstrom over the particle trajectories. The much higher efficiency in producing this type of ruptures for the neutrino recoil (O, 80 keV) is evident. It would be of interest to perform experiments like those of Refs. [17-20] with moderated ion beams that mimic the energy distributions of Fig. 1. To achieve this with present terrestrial neutrino sources seems more complicated, due to the limited fluxes and the sensitive dependence of T on the neutrino spectrum through $E_\nu^2$. This leads to a second remark: solar neutrinos are unable to produce the kind of damage discussed here. While the energy distribution of solar $^8$B neutrinos goes up to ~ 15 MeV, the recoils have lower mean energies and LET, and with a flux of $5.8 \cdot 10^6$ ν / cm² / s, it would take > 200 yr to induce the same recoils / kg as a stellar collapse at a distance as far as 100 pc. Neutrinos from the $^7$Be chain have a more competitive flux, but T < few tens of eV, which produce no important damage.

Neutrino recoils have the aggravating feature of inducing a uniformly-spread internal dose, able to reach highly radiosensitive tissue such as lymphatic cells or bone marrow. This is in contrast to high-LET components of the natural background like typical external alpha particles or protons, which are stopped in the outer layers of the skin. Only the fast neutron component of the secondary cosmic-ray spectrum is in principle able to induce similar recoils, albeit not uniformly distributed (10 cm of tissue reduce their relative dose by a factor $\gtrsim$ 10). The measured sea-level total dose induced by cosmic neutrons is $\sim 5 \cdot 10^{-7}$ Gy / year [21]. The nature of their recoils is however very different: more than 85% of first neutron interactions involve H atoms (this fraction becoming larger for ensuing recoils), with LET values below maximum RBE and the irreparability threshold, and an energy loss almost entirely due to ionization (Fig. 2). Heavier recoils (C,O) present a uniform distribution in their recoil energies up to the kinematically allowed maximum, diminishing their fraction in the T < 100 keV region, where RBE is maximal. Marine organisms have been markedly affected in several extinction episodes; the under-water environment is comparatively free of high-LET radiation, and collapse neutrinos would induce a totally unprecedented dose of it to sea creatures.

Progress in the incipient understanding of biological effects of low-dose, high-LET radiation will determine the importance of neutrino recoils. If the considerations presented are accurate, cataloguing of impending "neutrino bombs" in the galactic vicinity of the Sun may not be a superfluous task.

**acknowledgments**

I am indebted to the Groupe de Physique des Solides (U. Paris VII) for their hospitality during the time that this work was completed. I have profited from helpful coments from T. Girard. This work was supported in part by the National Science Foundation under Grant no. PHYS-9007847.

**Table 1.** Total recoil rate and average recoil energy <T> imparted by stellar collapse neutrinos (r= 10 pc) in some common constituents of organic matter. Also included are the projected range l of an ion of kinetic energy <T> in liquid water [11], its total stopping power $S_t$ and the fraction of it due to secondary atomic collisions, $S_n / S_t$.

| Element | R (recoils/kg) | <T> (keV) | l (μm) | $S_t$ (keV / μm) | $S_n / S_t$ | % weight in human body |
|---|---|---|---|---|---|---|
| H | $3.2 \cdot 10^{-2}$ | 152 | 2.0 | 82 | $1.22 \cdot 10^{-3}$ | 9.1 |
| C | $2.9 \cdot 10^{3}$ | 89 | 0.43 | 226 | 0.14 | 14.7 |
| N | $3.5 \cdot 10^{3}$ | 83 | 0.34 | 249 | 0.20 | 5.6 |
| O | $4.0 \cdot 10^{3}$ | 78 | 0.31 | 258 | 0.26 | 65.1 |
| P | $8.8 \cdot 10^{3}$ | 52 | 0.11 | 416 | 0.66 | 1.5 |
| Ca | $1.1 \cdot 10^{4}$ | 42 | 0.075 | 568 | 0.76 | 2.8 |

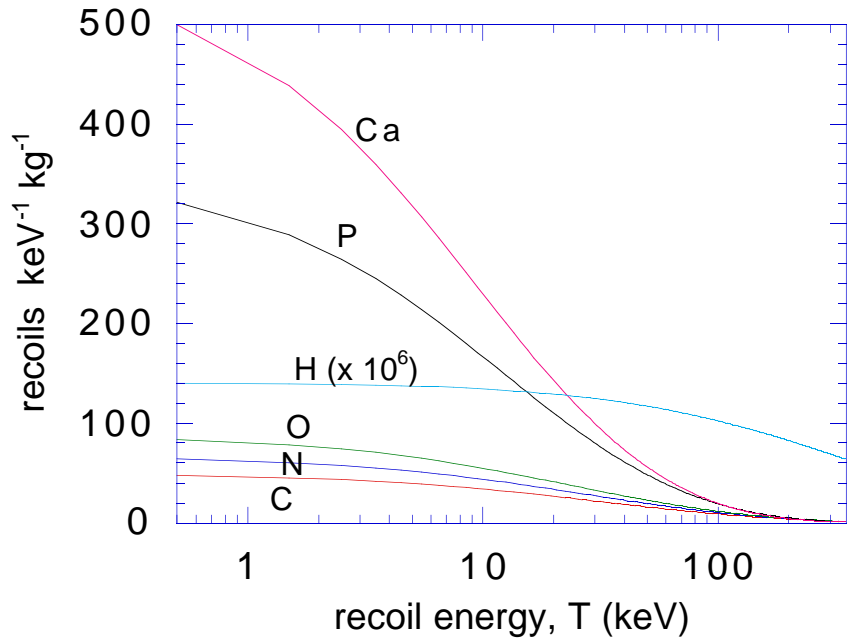

Fig. 1  The energy distribution of neutrino induced recoils in various constituents of organic tissue from a typical stellar collapse at a distance r = 10 pc (τ= 39.6 Myr). These differential rates scale with $r^{-2}$, as does the neutrino fluence F(r).

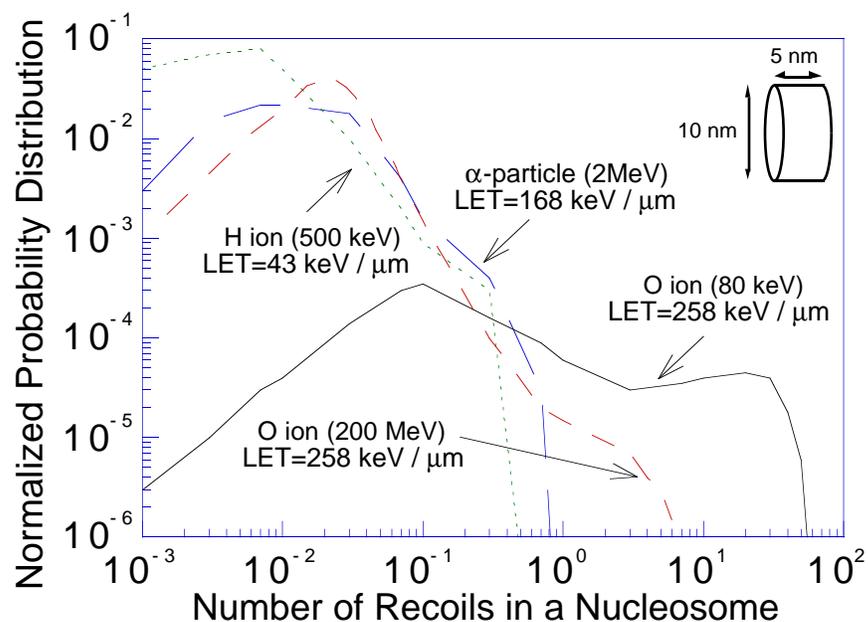

Fig. 2  Frequency of direct atomic displacements with recoil energy > 4.7 eV induced by the nuclear stopping power of different high-LET radiations in a water volume the size of a nucleosome. A 500 keV H ion is representative of the stopping of fission neutrons in tissue. Heavy ions can produce a significantly larger number of molecular dislocations through this mode of energy loss. The mean number of recoils is 0.034 (H), 0.26 ($\alpha$), 0.11 (O, 200 MeV) and 19.3 (O, 80 keV).